\documentclass[11pt,fullpage,doublespace,a4paper]{article}

\usepackage{cite}

\usepackage{epsfig}
\usepackage{epstopdf}
\usepackage[cmex10]{amsmath}
\usepackage{amssymb,mathrsfs}
\usepackage{bm}

\usepackage{subfigure}

\makeatletter
\newcommand*\rel@kern[1]{\kern#1\dimexpr\macc@kerna}
\newcommand*\widebar[1]{%
  \begingroup
  \def\mathaccent##1##2{%
    \rel@kern{0.8}%
    \overline{\rel@kern{-0.8}\macc@nucleus\rel@kern{0.2}}%
    \rel@kern{-0.2}%
  }%
  \macc@depth\@ne
  \let\math@bgroup\@empty \let\math@egroup\macc@set@skewchar
  \mathsurround\z@ \frozen@everymath{\mathgroup\macc@group\relax}%
  \macc@set@skewchar\relax
  \let\mathaccentV\macc@nested@a
  \macc@nested@a\relax111{#1}%
  \endgroup
}
\makeatother

\usepackage{url}

\usepackage{hyperref}

\usepackage[english]{babel}
\usepackage[pdftex]{color}
\usepackage{framed}
\newcommand{\comments}[1]{}

\usepackage{verbatim}

\usepackage{multirow}
\usepackage{lscape}
\usepackage{tabularx}

\usepackage[ruled]{algorithm2e}
\usepackage{pbox}
\usepackage{url}
\usepackage{epsfig}
\usepackage{listings}

\newcommand{\vx}{\mathbf x}

\newcommand{\oneb}{\mathbf 1}

\newcommand{\vs}{\mathbf s}

\newcommand{\vd}{\mathbf d}

\newcommand{\vf}{\mathbf f}
\newcommand{\vv}{\mathbf v}

\newcommand{\vt}{\mathbf t}

\newcommand{\vw}{\mathbf w}

\newcommand{\MD}{\mathbf D}

\newcommand{\MF}{\mathbf F}

\newcommand{\fabnote}[1]{\color{cyan}{#1 }\color{black}}
\newcommand{\bobnote}[1]{\color{blue}{#1 }\color{black}}
\newenvironment{tibenv}{\color{magenta}}{\color{black}}

\lstdefinestyle{tags}{
	formfeed=\newpage,
	basicstyle=\scriptsize\ttfamily,
	frame=single,
	numberstyle=\tiny,
	captionpos=b,
	breaklines=false
}

\begin{document}
%


\title{On Evaluation in Music Autotagging Research}


\title{On Local Generalization and Evaluation Validity in Music Autotagging Research}
\title{On Evaluation Validity in Music Autotagging}

%
%
%

\author{Fabien~Gouyon,
        Bob~L.~Sturm,
        Jo\~{a}o~Lobato~Oliveira,\\
        Nuno~Hespanhol,
        and~Thibault~Langlois}

\date{}
\maketitle

\begin{abstract}
Music autotagging, an established problem in Music Information Retrieval, aims to alleviate the human cost required to manually annotate collections of recorded music with textual labels by automating the process. Many autotagging systems have been proposed and evaluated by procedures and datasets that are now standard (used in MIREX, for instance). Very little work, however, has been dedicated to determine what these evaluations really mean about an autotagging system, or the comparison of two systems, for the problem of annotating music in the real world. In this article, we are concerned with explaining the figure of merit of an autotagging system evaluated with a standard approach. Specifically, does the figure of merit, or a comparison of figures of merit, warrant a conclusion about how well autotagging systems have learned to describe music with a specific vocabulary? The main contributions of this paper are a formalization of the notion of validity in autotagging evaluation, and a method to test it in general. We demonstrate the practical use of our method in experiments with three specific state-of-the-art autotagging systems --all of which are reproducible using the linked code and data. Our experiments show for these specific systems in a simple and objective two-class task that the standard evaluation approach does not provide valid indicators of their performance.
\end{abstract}

\maketitle

\section{Introduction}\label{intro}

Music autotagging is an established problem in Music Information Retrieval (MIR),
as witnessed by the publication of book chapters (e.g.,~\cite{Bertin-Mahieux2010c}), 
several journal articles (e.g.,~\cite{Turnbull2008,Bertin-Mahieux2008,Fu2011,Miotto2012})
and conference papers (e.g.,~\cite{Miotto2010b,Seyerlehner2010c,Xie2011,Marques2011b,Coviello2012,Nam2012b}),
PhD theses (e.g.,~\cite{Sordo2012}), tutorials (ISMIR 2013).
Music autotagging systems aim to annotate music audio signals with textual labels, or tags.
Ultimately, such systems could alleviate the human cost required to manually annotate collections of recorded 
music by automating the process.
Many music autotagging systems have been proposed and evaluated by procedures and datasets that are now standard,
as exemplified e.g. by six years of completed MIREX
``Audio Tag Classification'' task (ATC).
The topic of system evaluation itself plays a increasingly critical role in the MIR community,
as mentioned in the challenges highlighted in a recent Roadmap for MIR~\cite{roadmap2013}.

Clearly, the desire of this field of research is for an autotagging system, or any MIR system, to perform well in the real world.
%
One step towards considering how well MIR systems work in the real world is testing their robustness to a variety of environmental conditions, such as noise, audio quality, etc.
For instance, work has been dedicated to the effect of audio perturbations (e.g. adding white noise, filtering, different encodings, etc.) on the computation of 
low-level features such as MFCCs or chromas~\cite{Sigurdsson,JensenCEJ09,UrbanoISMIR2014},
and on the robustness to audio perturbations
of state-of-the-art systems for beat~tracking, chord recognition, and audio-to-score alignment~\cite{journals/taslp/GouyonKDATUC06,MauchE13}.

Whereas robustness tests seek to determine how sensitive a system is to characteristics of its environment, 
we contend the question that needs to be addressed first is whether a system's evaluation
provides us with valid conclusions about its true performance.
Indeed, virtually no autotagging 
evaluation has addressed the question of validity~\cite{UrbanoJIIS2013,Sturm2014a}. 

The main contributions of this paper are precisely a formalization of the notion of validity in autotagging evaluation, and a method to test it in general.
This method is based on the consideration that
if an autotagging system is pairing audio signals with tags in a meaningful way,
its behavior should not be significantly affected by irrelevant
perturbations of its input signals.
We perform several experiments demonstrating our method for three state-of-the-art autotagging systems.
We confirm in these experiments that the irrelevant perturbations we perform are ``fair'', i.e. they do not imply a significant 
covariate shift between the feature distributions of training and test data~\cite{Sugiyama2007,Quionero-Candela:2009:DSM:1462129}.


This article is organized as follows:
In the next section, we clarify the objectives of evaluation in music autotagging research,
review the standard approach to evaluation, and formalize the notion of validity in the context of
evaluation of autotagging systems.
Then, in Section~\ref{sec:testingevaluationR},
we present a method for testing the validity of autotagging evaluation,
based on specifically designed perturbations of test instances,
which we define as ``irrelevant transformations.''
Section~\ref{sec:experiments} describes our experiments
with this method in testing the validity of the evaluation
of three specific state-of-the-art autotagging systems.
We summarize the article and discuss its findings in Section~\ref{discussion}. 
All experiments and results in this article 
can be reproduced via data available on 
\url{http://www.fabiengouyon.org/}, under the ``Research'' -- ``Data for reproducible research'' menu item.


\section{Music Autotagging and its Evaluation}\label{sec:formalization}

\subsection{What is autotagging?}

Following 
\cite{Turnbull2008}, 
we consider music autotagging as a multi-label 
supervised learning problem with music audio signals as input, and
where the objective is to meaningfully relate tag concepts and acoustic phenomena.
Adopting the terminology of 
\cite{Seyerlehner2010c},
we equate music autotagging to ``transform[ing] an audio feature space into a semantic space,
where music is described by words'',
and we define a music autotagging system as one that 
annotates, i.e., assigns tags to, recorded music.
For example, if  singing voice is heard in the music, a good music autotagging system should annotate it with the tag ``vocals''.

\subsection{Current practices of music autotagging evaluation}\label{subsec:currentpractice}

An in-depth formalisation of evaluation in comparative experiments can be found in~\cite{Bailey},
and a preliminary application of it to the specific case of evaluation in MIR in~\cite{BobCMMR2014}.
%
A standard approach to music autotagging evaluation
is having a system annotate a set of signals,
and then comparing the resulting tags to the ``ground truth.''
Between 2008-2012, the MIREX\footnote{\url{http://www.music-ir.org/mirex/wiki/MIREX_HOME}}  
``Audio Tag Classification'' task (ATC) has employed this approach to 
systematically and rigorously evaluate
about 60 music autotagging solutions with standardized datasets.
This evaluation procedure also appears in many other works,
e.g.~\cite{Turnbull2008,Bertin-Mahieux2008,Miotto2010b,
Xie2011,Coviello2012,Nam2012b}.

A fundamental aspect of these evaluations is \emph{data}.
The music autotagging literature has established
a variety of benchmark datasets.
Several works use the datasets CAL500~\cite{Turnbull2008}, 
MagnaTagatune~\cite{Law2009},
and the Million Song Dataset~\cite{BertinMahieux2011}.
Among the datasets ATC uses
are MajorMiner~\cite{Mandel2008} and USPOP~\cite{Berenzweig2004}.
%
Evaluation in  music autotagging 
typically proceeds via cross-validation experiments,
as follows.
A dataset of sampled audio signals is partitioned
into $K$ non-overlapping folds.
This dataset is such that each signal is paired with ``ground truth'' tags from a given tag vocabulary.
Then, $K$ music autotagging systems are built
by training on the complement of a testing dataset fold.
The presence or absence of each tag from the ``ground truth'' 
is measured in the output of the system.
More specifically, the following \emph{measurements} are made: the number of true positives,
false positives, true negatives, and false negatives of each tag
are counted.

Music autotagging evaluation  involves computing several \emph{figures of merit} (FoM)
from these measurements.
In ATC, these include quantities named
``Average Tag Recall,'' ``Average Tag Precision,''
``Average Tag F-Measure,''
the precise meanings of which 
are specified in the source code of MIREX.\footnote{See method {\tt evaluateResultFold} in 
\url{https://code.google.com/p/nemadiy/source/browse/analytics/trunk/src/main/java/org/imirsel/nema/analytics/evaluation/tagsClassification/TagClassificationEvaluator.java}}
The ATC figure of merit ``Average Tag Recall'' 
is defined
as the mean of the $K$ micro-averaged recalls (also called ``global'' recalls);
the ``Average Tag Precision'' is defined as the mean of the
$K$ micro-averaged precisions;
and the ``Average Tag F-Measure'' is defined as
the mean harmonic mean of the $K$ ``Average Tag Precisions'' and ``Average Tag Recalls.''
Other figures of merit appear in the literature.
For instance, the macro-averaged recall of a system is defined as
the mean of the recalls of each tag.
This is also called per-tag recall \cite{Turnbull2008,Bertin-Mahieux2008,Miotto2010b,Marques2011b,Xie2011,Coviello2012,Nam2012b}.
Similarly, there is the macro-averaged precision, 
and macro-averaged F-measure.

\subsection{What can one expect from evaluating an autotagging system?}\label{subsec:whattoexpect}

Denote an autotagging system by $S$, which 
maps an input audio signal 
$\vx$ to a subset $\mathcal{X}$ of a set of tags, denoted $\mathcal{T}$. 
A~dataset is defined as an indexed set of tuples $(\vx, \mathcal{X})$. 
We notate the training dataset $\Psi$ and the testing dataset~$\Phi$.

A relatively common assumption to the design and evaluation of supervised learning systems,
such as autotagging systems,
is that the feature distributions of their training and test data are 
identical~(i.i.d.)~\cite{Quionero-Candela:2009:DSM:1462129}.
That is, that the features in $\Psi$ and $\Phi$ are sampled from the same distribution $\mathcal{D}$.
For instance, 
\cite{Marques2011b} illustrate the fact that state-of-the-art autotagging systems 
trained on a given dataset typically fail to generalize to datasets of different origins, where the i.i.d. assumption is not respected.
On the other hand, when the feature vectors of $\Psi$ and $\Phi$ are i.i.d., 
one should expect the performance of $S$ trained on $\Psi$ to be relatively stable with respect to
different sets $\Phi$.
This is  for instance the case when $\Psi$ and $\Phi$ are different folds (or combinations thereof) 
of the same dataset in a cross-validation procedure (see Section~\ref{subsec:currentpractice}).
One should therefore expect that $S$
 be put to use in ``similar conditions'' than those used for training.\footnote{
Note however that research in Domain Adaptation and Transfer Learning precisely address the design of systems 
coping with conditions different than those under which they were developed~\cite{Quionero-Candela:2009:DSM:1462129,PanY09TKDE,Ben-David,Sugiyama2007}.}


%

\subsection{Validity in music autotagging evaluation}\label{subsec:validity}

An evaluation of music autotagging systems produces measurements,
from which FoM are computed and conclusions then drawn.
For instance, when an FoM is significantly better for one system compared to another,
then one desires that the former system is better at autotagging  than the latter.
Hence, a critical question to answer is whether the approach used for evaluation is \emph{valid} for such conclusions,
i.e. whether ``we are really measuring what we want to measure''~\cite{UrbanoJIIS2013}.

More formally, denote by $\Gamma_{S}(t)$
the \emph{true performance} of a system $S$ on a tag $t \in \mathcal{T}$. 
(Note that $\Gamma_{S}(t)$ is a simplified notation for $\Gamma_{S;\Psi}(t)$,
as the system is a product of the training dataset $\Psi$.)
The true performance describes how well $S$ is expected to perform 
in using $t$ (or not) to annotate any test music audio signals
(assuming i.i.d. between train and test data).
Define $\Gamma_S(t)=\mathbb{E} \big[f_{S}(\vx,t)\big]$, 
where 
$\mathbb{E}\big[.\big]$ denotes the expectation 
over 
all possible feature vectors in the sample space,
and  $f_{S}(\vx,t)$ 
denotes some function that measures the discrepancy between the output of $S$ 
and whether $t$ truly applies to $\vx$
(e.g. if $f_{S}(\vx,t)$ is the $0/1-$loss,
$\Gamma_S(t)$ is the \emph{true risk}~\cite{Sugiyama2007}). 
Since we cannot evaluate this expectation (we do not have access to the true distribution of these features), 
$\Gamma_S(t)$ is not observable, and so it must be inferred from something observable.
Standard practice in music autotagging addresses this issue by evaluating $S$ on a 
test set $\Phi$,
and computing an \emph{estimated performance} $\widehat \Gamma_S(t)$ (e.g. \emph{empirical risk} in~\cite{Sugiyama2007}).
That is, computing a FoM on $\Phi$,
and inferring $\Gamma_S(t)$ from this.
(Note here again that $\widehat \Gamma_S(t)$ is a simplified notation for $\widehat \Gamma_{S;\Psi}(t,\Phi)$.)
This implicitly assumes that $\widehat \Gamma_S(t)$ and $\Gamma_S(t)$
are highly positively correlated.

We define an evaluation 
to be a {\em valid indicator of the true performance} $\Gamma_S(t)$
when:
\begin{equation}
[\widehat \Gamma_S(t) \textrm{ good}] \Leftrightarrow [\Gamma_S(t) \textrm{ high}]
\label{eq:ass1}
\end{equation}
and when, for two systems $S_1$, $S_2$
\begin{equation}
[\widehat \Gamma_{S_1}(t) \textrm{ better than } \widehat \Gamma_{S_2}(t) ] \Leftrightarrow [\Gamma_{S_1}(t) \textrm{ higher than } \Gamma_{S_2}(t)]
\label{eq:ass2}
\end{equation}
where $\Leftrightarrow$ is logical equivalence.
In other words, (\ref{eq:ass1}) says a valid 
evaluation of $S$ produces a good FoM on $t$ 
if and only if the true performance of $S$ on $t$ is indeed high;
and (\ref{eq:ass2}) says a valid evaluation 
produces a better figure of merit for $S_1$ than for $S_2$ on $t$
if and only if the true performance of $S_1$ is higher than that of  $S_2$ on $t$.

If, for an evaluation making use of a test set $\Phi$, (\ref{eq:ass1}) and~(\ref{eq:ass2}) do not hold for some tag $t$,
then that evaluation is not a valid indicator of the true performance of $S$ 
on $t$.
The principal question is no longer,
``How good/bad is $\widehat \Gamma_S(t)$?'',
or, ``Is $\widehat \Gamma_{S_1}(t)$ significantly higher/lower than $\widehat \Gamma_{S_2}(t)$?'',
but now, ``Does the evaluation of $S$ in $\Phi$ provide a valid indication of its
true performance on $t$?''

\section{A method for testing evaluation validity}\label{sec:testingevaluationR}

According to the notion of validity defined in Section~\ref{subsec:validity},
we now present a method for testing the validity of  
the evaluation of music autotagging systems.
%
The basic rationale is the following: In experimental conditions where one should expect the
true performance of an autotagging system to be relatively stable (see Section~\ref{subsec:whattoexpect}),
if its estimated performance varies such that (\ref{eq:ass1}) and (\ref{eq:ass2}) are violated, then
that evaluation is not a valid indicator of the system's true performance.

At its core, our method is based on a systematic search for 
perceptually indistinguishable test sets, while controlling for the
required absence of covariate shift~\cite{Sugiyama2007,Quionero-Candela:2009:DSM:1462129}.
These test sets are obtained by irrelevant transformations of a limited selection of instances
in a test set. Our approach is comparable to that of 
\cite{Szegedy2014}, who
 test the local generalization capability of their image classification systems. 
Szegedy et al. show, on three different benchmark datasets (images in their case),
that for every test instance that is correctly classified by any of the state-of-the-art 
systems
they studied (deep neural networks), 
there exists instances in the local vicinity of the original test instance that are perceptually indistinguishable 
from the original but that are misclassified by the 
system, in any of the possible classes.
They obtain these ``adversarial'' instances (which they also refer to as ``blind spots'') 
by means of ``imperceptible'' transformations of test instances, found by optimizing the input to
maximize the prediction error, while restricting the optimization
process to local space around the original test instance.
While Szegedy et al. employ a constrained optimization approach to find these adversarial instances, 
we use a brute force approach to achieve the same results. Furthermore, our aim is not to show the 
existence of ``blind spots'', but of testing (\ref{eq:ass1}) and (\ref{eq:ass2}) for a system.
 
\subsection{Our method}

More formally, consider \(\mathcal{T} = \{t, \bar t\}$,
where $\bar t$ is the negation of $t$.
For a $S$, assume $\Gamma_S(t)$ and $\Gamma_S(\bar t)$ remain constant, 
i.e., $S$ does not learn about $\mathcal{T}$ after its initial training.
Consider a testing dataset $\Phi$ of audio signals,
each tagged $t$ or $\bar t$.
Define the transformation of the testing dataset, 
$\mathcal{F}(\Phi) = \{(F_i(\vx_i), \mathcal{X}_i) : i \in \mathcal{I}\}$,
where $F_i$ transforms the audio signal~$\vx_i$, and $\mathcal{I}$ denotes the set of indexes of $\Phi$.
Adapting the notion proposed in~\cite{Sturm2014a},
we define $\mathcal{F}(\Phi)$ as an \emph{irrelevant transformation} of $\Phi$
if it complies with the following requirements:
\begin{itemize}
\item $\forall F_i(\vx_i)$, $\vx_i$ and $F_i(\vx_i)$ are perceptually indistinguishable,
i.e., a human describing $\vx_i$ as $t$ will also describe $F_i(\vx_i)$ as $t$.
\item $\mathcal{F}(\Phi)$ produces no covariate shift with respect to $\Phi$~\cite{Sugiyama2007,Quionero-Candela:2009:DSM:1462129}.
\end{itemize}

\begin{figure}[h!]
\centering
\includegraphics[width=\columnwidth/5*4]{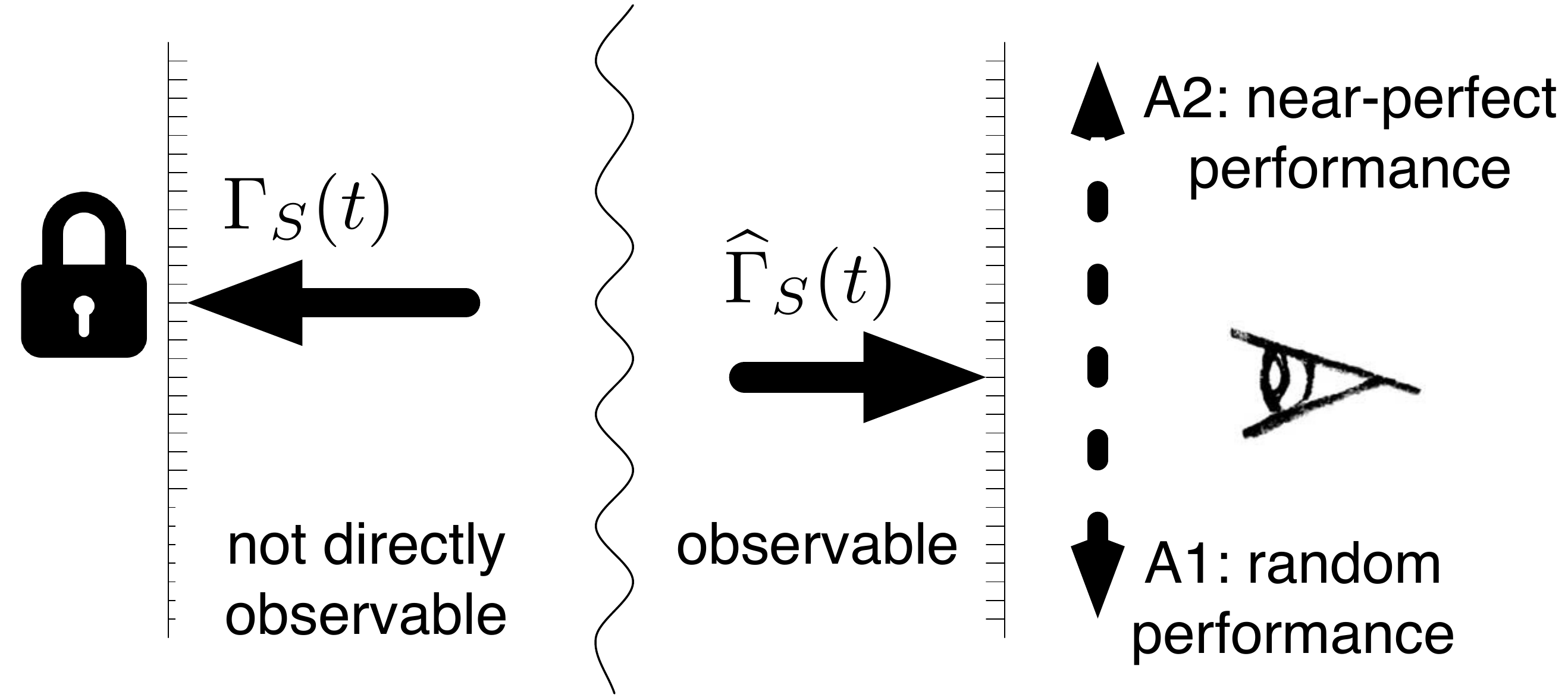}
\caption{
To prove (\ref{eq:ass1}) does not hold,
while the true performance of $S$ on $t$, $\Gamma_S(t)$,
remains constant (whatever its value),
we devise experimental conditions so that
its estimator, the figure of merit $\widehat \Gamma_S(t)$ takes values
ranging from 
random to close to perfect.
}
\label{fig:twohands}
\vspace{-0.1in}
\end{figure}

Consider $\widehat \Gamma_S(t)$ is significantly better than random.
With regards to (\ref{eq:ass1}), 
we thus attempt the following tasks:
\begin{enumerate}
\item[A1.] Find $\mathcal{F}$ to transform $\Phi$ 
such that $\widehat \Gamma_S(t)$ is not significantly better than random.
\item[A2.] Find $\mathcal{F}$ to transform $\Phi$ such that 
$\widehat \Gamma_S(t)$ is close to perfect.
\end{enumerate}
If we can accomplish A1 and A2, (\ref{eq:ass1}) does not hold
because $\widehat \Gamma_S(t)$ 
can change between extremes though $\Gamma_S(t)$ stays the same.
Procedures A1 and A2 are schematized in figure~\ref{fig:twohands}.

Now, 
with regards to (\ref{eq:ass2}),
given two 
systems
$S_1$ and $S_2$, 
we attempt the following:
\begin{enumerate}
\item[B1.] Find $\mathcal{F}$ to transform $\Phi$ 
such that $\widehat \Gamma_{S_1}(t)$ 
is significantly better than $\widehat \Gamma_{S_2}(t)$.
\item[B2.] Find $\mathcal{F}$ to transform $\Phi$ 
such that $\widehat \Gamma_{S_2}(t)$ 
is significantly better than $\widehat \Gamma_{S_1}(t)$.
\end{enumerate}
If we can accomplish B1 and B2, (\ref{eq:ass2}) does not hold
because we can make the relative figures of merit of two systems 
significantly different in either direction while their relative 
true performance, and ranking, does not change.

\subsection{Statistical significance}
Task A1 essentially attempts to make the performance
of $S$ on \(\Phi\) decay to the point that it is no longer inconsistent with
that of a random system.
We thus analyze the behavior of a system that 
independently picks $t$ for an input with probability $p_t$
(and~$\bar t$ with probability $1-p_t$).
Denote this system by $R(p_t)$.
%
%
Of the $N$ signals in $\Phi$,
consider that there are $n_t$ tagged with $t$,
and $n_{\bar t}$ tagged with $\bar t$.
Let $X$ and $Y$ be random variables for the number of 
{\em correct} tags by $R(p_t)$ of $t$ signals
and $\bar t$ signals, respectively.
The probability of $X=x$ 
is distributed $X \sim Bin(n_t, p_t)$;
and of $Y=y$ is distributed $Y \sim Bin (n_{\bar t}, 1-p_t)$.
The joint probability of \{$X=x$, $Y=y$\} is thus:
\begin{align}
P_{X,Y}(x,y; p_t) & = {n_t \choose x}p_t^x(1-p_t)^{n_t-x}{n_{\bar t} \choose y}(1-p_t)^yp_t^{n_{\bar t}-y}
\end{align}
for $0 \le x \le n_t$, $0 \le y \le n_{\bar t}$,
and zero elsewhere.

Now, consider $S$ produces \(\{x,y\}\) in $\Phi$.
For A1, we test the null hypothesis $H_{0A_1}$: 
results at least as good as \(\{x,y\}\)
are expected from an element of $\{R(p_t): p_t \in [0,1]\}$.
In other words, observations at least as good as \(\{x,y\}\) 
are 
consistent
with what we expect to be produced by a random system.
We test $H_{0A_1}$ by computing:
\begin{equation}
\max_{p_t \in [0,1]} P[X \ge x, Y \ge y; p_t] =
\max_{p_t \in [0,1]}\sum_{i=x}^{n_t}\sum_{j=y}^{n_{\bar t}}P_{X,Y}(i,j; p_t). 
\label{eq:pvalueexp1}
\end{equation}
and fail to reject $H_{0A_1}$ when this value is greater than 
the statistical significance parameter $\alpha$.
Recall that our goal with A1 is to show that 
$\mathcal{F}(\Phi)$ leads to a failure to reject $H_{0A_1}$
though we can reject it for $\Phi$.

For B1 and B2, we must compare the performance of two systems on the same dataset.
We count the total number of signals $b$ for which 
$S_1$ and $S_2$ contradict each other, i.e. only one of the systems is wrong.
Denote $a_{12}$ the number
of signals in the dataset where $S_1$ makes correct predictions and $S_2$ is wrong ($b=a_{12}+a_{21}$).
If either system is equally likely to be correct (i.e. $a_{12}$ should not be significantly different from $a_{21}$),
then we expect $a_{12}$ to not be significantly different from $b/2$.
For B1, the null hypothesis $H_{0B_1}$ is thus $a_{12} = b/2$.
Define the random variable $A_{12} \sim Bin(b, 0.5)$
to model $a_{12}$ in $b$ independent trials
when $S_1$ and $S_2$ are equally likely to be
correct when they contradict each other.

Given an observation for $a_{12}$, we compute the probability that $A_{12}$ is at least as large as $a_{12}$
as:
\begin{equation}
P[A_{12} \ge a_{12}] =
\sum_{x=a_{12}}^{b} {b \choose x} 0.5^b.
\label{eq:pvalueexpB1}
\end{equation}
If $P[A_{12} \ge a_{12}] < \alpha$, then we reject $H_{0B_1}$.
We follow the same reasoning for B2, and if $P[A_{21} \ge a_{21}] < \alpha$, then we reject $H_{0B_2}$.


\section{Experiments}\label{sec:experiments}

Here, we first detail our methodology 
for applying in practice the method 
defined in Section~\ref{sec:testingevaluationR} for evaluating three state-of-the-art systems with three standard datasets.
We then present evidence of the irrelevance of the transformations 
in our experiments.
We finally present results on absolute and relative performance
of the tested systems, showing that their evaluations are not valid indicators of true performance.
In other words, they
do not provide valid indicators for concluding whether any of them is objectively good, or better than any other.

\subsection{Methodology}\label{sec:Methodology}
We test (\ref{eq:ass1}) and (\ref{eq:ass2}) for all systems resulting from 
three state-of-the-art music autotagging approaches crossed with folds of
three datasets commonly used for evaluation in music autotagging.
We set $t$ as the tag ``Vocals'', 
i.e., whether a piece of music includes singing voice or not.
We justify this choice by the fact that compared to other possible tags,
the tags ``Vocals'' ($t$) and ``Non-Vocals'' ($\bar t$) are better defined 
and more objective relative to other kinds of tags, e.g., genre and emotion,
and that it appears in all of our three datasets in some form, 
e.g., ``voice'', ``gravely voice'', or ``female singer''.
This scenario is simpler than the general case of autotagging,
but we claim that if the evaluation of a given system
can be shown not to provide a valid indication of true performance
for such an objective, single-label case, 
it is not reasonable to assume that the evaluation of that system
should be valid in the more subjective and ill-defined general multilabel case (we discuss this further in Section~\ref{discussion}).
%
%
It should also be noted that not only is such a tag suitable to the experimental procedure in this article,
but also the actual ability to automatically detect whether a music excerpt includes singing voice or not 
corresponds to a realistic and very useful problem.

\subsubsection{Deflation and inflation procedures}

Given a system $S$ and test dataset $\Phi$,
we test (\ref{eq:ass1}) using what we call ``deflation'' and ``inflation'' procedures,
that are illustrated in
Algorithms~\ref{deflation} and~\ref{inflation} (where $I\vx = \vx$ is the identity transformation). 
%
%
For deflation, we find irrelevant transformations $\mathcal{F}(\Phi)$
that decrease the number of correct responses by $S$.
As mentioned in Section~\ref{sec:testingevaluationR}, this is comparable to the procedure of 
\cite{Szegedy2014}
(in the context of image classification) where for each possible test instance correctly classified by 
a system they find in its local vicinity an ``adversarial'' instance that is misclassified,
although they are perceptually indistinguishable. 
%
In the deflation procedure, we alternate between finding elements of $\Phi$ for which
$S$ is correct, and transforming these signals 
in irrelevant ways (as defined in Section~\ref{sec:testingevaluationR})
to make $S$ respond incorrectly, until the performance of $S$ becomes similar
to that of a random system, according to (\ref{eq:pvalueexp1}) (with $\alpha=0.01$). 
For inflation, we find transformations $\mathcal{F}(\Phi)$ 
that increase the number of correct responses by $S$.
To do this, we alternate between finding elements of $\Phi$ for which
$S$ is incorrect, and transforming these signals
in irrelevant ways to make $S$ respond correctly.
The system's true performance $\Gamma_{S}(t)$ never changes,
but the deflation procedure attempts to make its FoM $\widehat \Gamma_S(t)$ worse,
while the inflation procedure attempts to make it better.
(Note that in both procedures a given signal 
is transformed at most once and that we seek to transform only a few instances in $\Phi$.)
%
If we are able to produce any FoM of a system
just by changing irrelevant aspects of $\Phi$ 
(i.e. transformations do not produce a covariate shift and
are perceptually indistinguishable), then (\ref{eq:ass1}) does not hold.

\begin{algorithm}[t] \footnotesize
\SetAlgoLined
\SetKw{Init}{Initialization:}
\Init
\Begin{
\nl $\mathcal{F} \leftarrow  \{F_i = I : i \in \mathcal{I}\}$ (Initialize all transformations to identity)\;
}
\Repeat{the figure of merit of $S$ on the transformed dataset is no better than random}{
\nl $\mathcal{J} \leftarrow \{i \in \mathcal{I}: F_i \in \mathcal{F}\left (S(F_i\vx_i) = \mathcal{T}_i \right) \}$ (indices of signals for which $S$ produces correct tags)\;
\nl Produce irrelevant transformation, $G$\;
\nl $\mathcal{F} \leftarrow \{F_i = G : i \in \mathcal{J}\} \bigcup \{F_i \in \mathcal{F} : i \in \mathcal{I} \backslash \mathcal{J}\}$ (update set of transformations)\;
}
    \caption{Pseudo-code for the deflation procedure.}\label{deflation}
    
\end{algorithm}

\begin{algorithm}[t] \footnotesize
\SetAlgoLined
\SetKw{Init}{Initialization:}
\Init
\Begin{
\nl $\mathcal{F} \leftarrow  \{F_i = I : i \in \mathcal{I}\}$ (Initialize all transformations to identity)\;
}
\Repeat{the figure of merit of $S$ on the transformed dataset is close to perfect}{
\nl $\mathcal{J} \leftarrow \{i \in \mathcal{I}: F_i \in \mathcal{F}\left (S(F_i\vx_i) \ne \mathcal{T}_i \right) \}$ (indices of signals for which $S$ produces incorrect tags)\;
\nl Produce irrelevant transformation, $G$\;
\nl $\mathcal{F} \leftarrow \{F_i = G : i \in \mathcal{J}\} \bigcup \{F_i \in \mathcal{F} : i \in \mathcal{I} \backslash \mathcal{J}\}$ (update set of transformations)\;
}
    \caption{Pseudo-code for the inflation procedure.}\label{inflation}
    
\end{algorithm}

We test (\ref{eq:ass2}) using the same iterative procedure, 
but with two systems.
Given $S_1,S_2$ and $\Phi$,
we set aside all instances of $\Phi$ for which $S_1$ is correct, but $S_2$ is not.
Then we apply successive transformations to the remaining instances until the performance of $S_1$
becomes significantly better than that of $S_2$, according to (\ref{eq:pvalueexpB1}) (with $\alpha=0.01$).
We repeat this procedure, but set aside all instances of $\Phi$ for which $S_2$ is correct
and $S_1$ not,
then we apply successive transformations to the remaining instances until the performance of $S_2$ becomes significantly better than that of $S_1$.

\subsubsection{Signal transformations}\label{transform}
\begin{figure}[h]
\centering
\includegraphics[width=10cm]{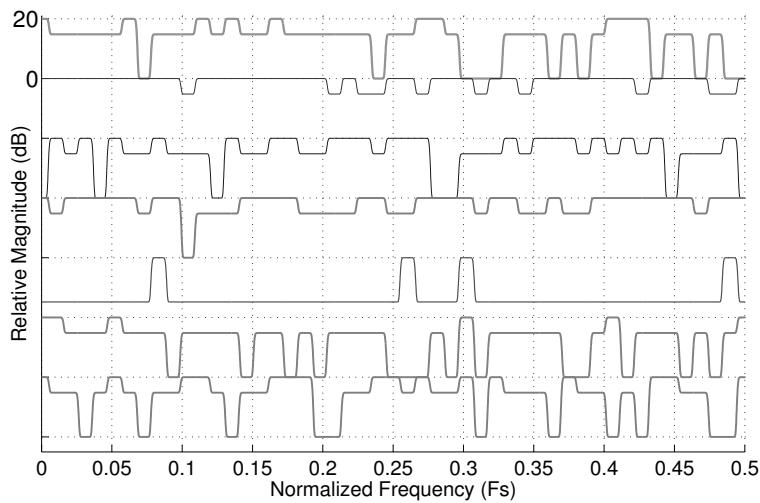}
\caption{ Magnitude responses of a selection of filters 
used in 
the deflation procedure. 
Note that the y-axis is ``relative magnitude''.}
\label{fig:filters}
\vspace{-0.1in}
\end{figure}

Our method in Section~\ref{sec:testingevaluationR} does not specify the nature of the irrelevant transformation. 
This depends on the tag. In our case for Vocals/Non-Vocals tags, examples of transformations that would not be 
irrelevant are e.g. adding voice to signals without voice, and removing vocals from signals that have voice. 
Examples of irrelevant transformations for Vocals/Non-Vocals tags may be minor time-stretching and/or pitch-shifting, 
changes in instrumentation while preserving voice or no voice, minor equalization, and so on. 
In our experiments here, we use time-invariant filtering, which proceeds as follows. 
We use the same irrelevant transformation, as well as time-stretching in another work~\cite{Sturm2014}: 
Specifically, we first build a 96-channel near perfect reconstruction polyphase filterbank.\footnote{
We adopt this code: \url{http://www.mathworks.com/matlabcentral/fileexchange/15813-near-perfect-reconstruction-polyphase-filterbank}}
Passing a signal through this filterbank 
produces 96 signals
that when added with unity gain reproduces the original signal
with an average reconstruction squared error of -300~dB.
We, however, reduce the gains of a randomly selected subset of the 96 channels
and then sum the outputs of the filterbank.
This subset can be any number of channels,
and the attenuation of each channel selected 
is bounded to be no more than 20~dB.
This results in numerous different filters that ``equalize'' 
audio signals but preserve the music they embody.
Figure~\ref{fig:filters} shows
the magnitude responses of some of these filters.
In Section~\ref{irrelevance}, we test
the irrelevance of these transformations.
Audio examples and software code are available 
on the article's companion webpage (which link is provided in Section~\ref{intro}).


\subsection{Data}\label{data}

We now discuss the data we use, and our preprocessing of it.
Table~\ref{tab:stats} provides data statistics.
Data folds are available on the article's companion webpage (link in in Section~\ref{intro}).
We use three different datasets, 
CAL500, a subset of MagnaTagatune, 
and a subset of the Million Song Dataset,
each described below.
We reduce the vocabulary of each dataset to the Vocals and Non-Vocals tags,
i.e. we keep all instances annotated with a tag corresponding to either Vocals or Non-Vocals tags,
we do not consider further the remaining instances.
In this process, we favor data \emph{quality} over \emph{coverage}, this has the advantage
to make exhaustive listening and checking feasible, offering hence the guarantee
of data with no noise in annotations.
We correct annotations of the resulting data via a careful listening.
The tags Vocals and Non-Vocals are 
well-defined and relatively objective,
mutually exclusive, and 
always relevant.
It is thus straightforward to manually
clean and correct annotations of our three datasets with respect to these tags.
We split each dataset into folds,
and artist filtering~\cite{Pampalk2005,Flexer2007} is used to guarantee that no same artist appears in both training and test data.

\begin{table}[h]
\centering
\begin{tabular}{l|c|c|c}
& CAL500 & MagTag5k & MSD24k \\
\hline
\hline
 Vocals pieces & 444                 & 1626        & 1146            \\
 Non-Vocals pieces  & 58                 & 723         & 531              \\
\hline
Total & 502 & 2349 & 1677 \\
\end{tabular}
\caption{Statistics for the datasets used in  experiments.}
\label{tab:stats}
\end{table}

\subsubsection{CAL500}

This dataset is a collection of 502 music pieces 
annotated from a vocabulary of 174 tags.
It was first introduced in~\cite{Turnbull2008},
it is available online
and is widely used in the autotagging literature.
When obtained from the original website,
we found that all sound files but two were there, although their annotations were.
Thus, we corrected this by retrieving the missing songs.

We consider songs originally annotated with tags such as ``Female lead vocals'', or ``Vocals-Gravelly'' 
instances of the Vocals tag (see the full list in Appendix~A).
There is no explicit Non-Vocals tags in CAL500, so
we initially considered all remaining songs as instances of the Non-Vocals tag,
and after careful listening, retagged 11 instances from Non-Vocals to Vocals.
The dataset is  divided in 2 folds.\footnote{
We chose 2 folds and not 3 (as with the other datasets) because of the relative few  Non-Vocals instances (58) in the whole dataset.}


\subsubsection{MagTag5k}
This is a processed version of the original 
MagnaTagatune
dataset
(originally 21,642 pieces and a vocabulary of 188 tags~\cite{Law2009}),
coping for issues of duplication, synonymy, etc., in the original dataset. 
Details about the preprocessing applied on that dataset can be found in~\cite{Marques2011b}.
This dataset consists of 5,259 music pieces annotated from a vocabulary of 137 tags,
and is available online.\footnote{\url{http://tl.di.fc.ul.pt/t/magtag5k.zip}}

We assign the Vocals tag to songs annotated with the tags ``female.singing'', ``male.singing'', or ``singing''.
We assign the Non-Vocals tag to songs annotated with the tags ``no.singing''.
This yields 2,393 songs, which we check by careful listening, after
which the final dataset contains 2,349 instances, see Table~\ref{tab:stats}.
The dataset is divided in 3 folds.


\subsubsection{MSD24k}

We designed the MSD24k dataset for in-house experiments in music autotagging,
with the main objective to set up a dataset, comprising the audio data,
with tags of relatively good quality and with the highest density of annotations 
possible (i.e. imposing a lower limit on the number of tags per music piece).
As this article is the first publication referring to it, we now describe the procedure
followed in its creation.

This dataset is based upon the subset of the  Million Song 
Dataset (MSD)~\cite{BertinMahieux2011}
for which the MSD website provides\footnote{on \url{http://labrosa.ee.columbia.edu/millionsong/lastfm}}  
{\tt Last.fm} tags associated to its tracks (943,347 tracks).  
In order to cope with the significant problem of noise in {\tt Last.fm} tags~\cite{Lamere2008},
we follow the same rationale as~\cite{Tingle2010}
and focus on tags with clear musical meaning, as defined by teams of musicologists
of the Music Genome Project at the Pandora Internet radio.
We therefore generate a \emph{relevant tag vocabulary}~$\mathcal{T}$ consisting
of the overlap between Pandora tags (gathered from the CAL10k dataset~\cite{Tingle2010})
and existing {\tt Last.fm} tags from MSD. This vocabulary contains 708 tags.
Retrieving the music pieces from MSD with at least 1 tag in $\mathcal{T}$ yields a total of
257,387 pieces.
We then keep only pieces with \emph{at least 4 tags per piece}, lowering the total number of pieces to 60,769.
Of these, we were only able to retrieve 30~s snippets of 36,000 pieces in mp3 format.
Removing duplicates yields 26,277 pieces.
We finally remove the pieces corresponding to the ``list of MSD \{song ID, track ID\} pairs that should not be 
trusted'' (list available 
online).\footnote{\url{http://labrosa.ee.columbia.edu/millionsong/sites/default/files/tasteprofile/sid_mismatches.txt}}
This yields a final amount of 23,740 music pieces annotated from a vocabulary of 265 tags.

We assign the Vocals tag to songs annotated with tags such as
``A breathy male lead vocalist'', or ``A distinctive male lead vocalist''. 
Appendix~A 
lists the full tag list.
As for the CAL500 dataset, there is no explicit Non-Vocals tags in MSD24k, however
in that case the  dataset size makes very difficult an exhaustive listening. Therefore,
we recur to the following heuristics to select Non-Vocals instances.
We divide the dataset in 2 groups: Group~A made up of songs in the Vocals tag,
and Group~B made up of the remainder.
We then rank all tags according to their representativeness of both groups, from
``occuring mostly in songs from Group~A'', to ``occuring mostly in songs from Group~B''.
We then take a random sample of 1000 songs annotated only with the most representative tags of Group~B.
After careful listening to these songs, we keep 531 instances of the Non-Vocals tag.
(Note here that with this procedure, we favor quality over coverage of Non-Vocals instances.)
The dataset is divided in 3 folds.


\subsection{Building Music Autotagging Systems}\label{systems}
We use three different approaches to build music autotagging systems.
The first, SVMBFFs, combines bags of frames of features (BFFs)
and a support vector machine classifier (SVM).
The second, VQMM, first codes a signal using vector quantization (VQ)
in a learned codebook, and then estimates
conditional probabilities in first-order Markov models (MM).
The third, SRCAM, employs sparse representation classification
to approximate a high-dimensional 
psychoacoustically-motivated frequency modulation feature.
Below, we discuss each approach in more detail.

\subsubsection{SVMBFFs}\label{marsyas}
This approach, a variant of one proposed by 
\cite{Ness2009},
trains a linear SVM to output probabilities from an input BFFs,
from which tags are selected.
The BFFs, which are 68-dimensional vectors, 
are means and standard deviations computed from
texture windows of 30 s of analysis frames of 23.2 ms duration
(and overlapped by 50\%).
The 17 low-level features extracted from each frame are:
zero crossing rate, spectral centroid, roll-off and flux, 
and the first 13 
mel-frequency cepstral coefficients (MFCCs).
SVMBFFs trains an SVM by a ``normalized'' training dataset of BFFs,
i.e., where each dimension of the set of transformed BFFs lies in \([0,1]\).
We use the SVMBFFs implementation available in
the MARSYAS framework.\footnote{MARSYAS can be downloaded here: \url{http://marsyas.info/}. 
We use default settings of \texttt{bextract} 
and \texttt{kea} v.5099.}

\subsubsection{VQMM}
This approach computes the 13 MFCCs after the zeroth
with an analysis frame of 93 ms using the YAAFE toolbox.\footnote{\url{http://yaafe.sourceforge.net/}}
Analysis frames are overlapped by 50\%.
Given the 
feature vectors \(\{\vf_1, \vf_2 \ldots, \vf_n\}\) extracted from an input signal, 
VQMM first expresses it as an ordered code \(\{w_1, w_2, \ldots, w_n\}\) in a codebook \(\mathcal{C}\),
then computes a probability of observing this code in each of 
a set of duples of models \(\{ (M_t, \widebar{M}_t) : t \in \mathcal{T}\}\),
and finally selects a set of tags from $\mathcal{T}$ based on maximum likelihood.
The duple of models \((M_t, \widebar{M}_t)\) is composed
of a model $M_t$ trained on coded features for which
the tag $t \in \mathcal{T}$ is relevant,
and a model $\widebar{M}_t$ trained on coded features for which it is not relevant.
In our case, $M_t$ models ``Vocals'', 
and $\widebar{M}_t$ models ``Non-Vocals''.
VQMM computes the probability of observing the ordered code \(\{w_1, w_2, \ldots, w_n\}\) 
in the model of tag $t \in \mathcal{T}$,
$P_{M_t}(w_1, w_2, \ldots, w_n)$,
as well as its complement, $P_{\widebar{M}_t}(w_1, w_2, \ldots, w_n)$.
If $P_{M_t}(w_1, w_2, \ldots, w_n) > P_{\widebar{M}_t}(w_1, w_2, \ldots, w_n)$,
VQMM selects $t$ as a tag for the input.

VQMM builds a codebook by first grouping all features extracted from the signals in a training dataset
into $K=75$~clusters using $k$-means \cite{Gersho1991} --though other unsupervised approaches could be used-- 
and then pairing the \(K\) centroids of the clusters with codewords.
To code a feature vector in terms of the codebook,
VQMM selects the codeword of the nearest (in a Euclidean sense)
centroid in the codebook.

VQMM builds a model under the assumption that
the ordered code is a first-order Markov process, i.e.,
all pairs of elements from an ordered code \(\{w_1, w_2, \ldots, w_n\}\),
except for those that are subsequent, are independent.
The log joint probability of this code in ${M}_t$ thus becomes
\begin{equation}
\log P_{M_t}(w_1, w_2, \ldots, w_n) = \log P_{M_t}(w_1) + \sum_{i=1}^{n-1} \log P_{M_t}(w_{i+1}|w_i).
\end{equation}
VQMM trains ${M}_t$
by estimating the set of conditional probabilities \(\{P_{M_t}(w_i|w_j) : w_i, w_j \in \mathcal{C}\}\),
as well as \(\{P_{M_t}(w_i) : w_i \in \mathcal{C}\}\),
from coded feature vectors extracted from the 
training instances for which \(t\) is a relevant tag.
VQMM uses the coded features of all other signals 
to train  $\widebar{M}_t$.
More details can be found in~\cite{langlois2009music}.\footnote{Source code is available
at \url{https://bitbucket.org/ThibaultLanglois/vqmm}.}

\subsubsection{SRCAM}
This approach, a variant of one proposed by 
\cite{Panagakis2009,Sturm2012}
and 
\cite{Sturm2012c},
uses sparse representation classification (SRC) \cite{Wright2009b}
of auditory temporal modulation features (AM).
Here, we extend it to a multilabel classifier.
Given the {\em dictionary} of feature atom-tag atom duples
\(  \{(\vd_i,\vt_i/\|\vt_i\|_2) : i \in \mathcal{I}\}\),
SRCAM approximates a feature vector $\vf$ 
as a linear combination of a small number of feature atoms,
and then produces a tag vector $\vt	$ by thresholding a linear combination of the tag atoms.

More formally, SRCAM first solves
\begin{equation}
\min_\vs \|\vs\|_1 \; \textrm{subject to} \; \left \| \frac{\vf}{\|\vf\|_2} - [\vd_1|\vd_2|\cdots] \vs \right \|_2^2 \le \epsilon^2
\label{eq:SRCAMl1}
\end{equation}
then uses the solution $\vs$ to produce
the linear combination of tag atoms
$\vw = [\vt_1/\|\vt_1\|_2 \, | \, \vt_2/\|\vt_2\|_2 \, |\cdots ]\vs$,
and finally produces from this the tag vector $\vt = T_\lambda(\vw/\|\vw\|_\infty)$,
where $T_\lambda(\cdot)$ is a threshold operator,
its $i$th element defined 
\begin{equation}
[T_\lambda(\vw/\|\vw\|_\infty)]_i =
\begin{cases}
1, & [\vw]_i/\|\vw\|_\infty > \lambda \\
0, & \textrm{else}.
\end{cases}
\label{eq:SRCAMthreshold}
\end{equation}
The non-zero dimensions of $\vt$ correspond to the 
tags in $\mathcal{V}$ considered relevant for 
annotating
the input signal.

SRCAM defines the dictionary from a training feature-tag vector dataset
by first constructing a matrix of the features, $\MF = [\vf_1|\vf_2|\ldots]$,
finding the maximum and minimum of each dimension,
defined as column vectors $\max\MF$ and $\min\MF$, respectively, 
and then computing the matrix of {\em normalized} feature atoms
\begin{equation}
\MD = [\vd_1 |Ê\vd_2 |\cdots] = \textrm{diag}(\max\MF-\min\MF) (\MF - \oneb [\min\MF]^T).
\end{equation}
Normalization guarantees that each dimension of $\MD$ is in $[0,1]$.

The particulars of our implementation of SRCAM are as follows.
We solve (\ref{eq:SRCAMl1}) using SPGL1~\cite{Berg2008}, 
and define $\epsilon^2 = 0.01$ and 200 iterations from experimentation.
For thresholding (\ref{eq:SRCAMthreshold}),
we define $\lambda = 0.25$ from experimentation.
We compute features from contiguous segments of about 27.7 s duration in a signal.
Specifics about computing AMs
are given in \cite{Sturm2012c}.

\subsubsection{Baseline Results}
We test these systems on the CAL500 dataset,
but restricted to the 97 most frequent tags (as done
in~\cite{Miotto2010b,Xie2011,Nam2012b,Coviello2012}).
We use 5-fold cross-validation,
and compute (as is standard in autotagging research) 
the mean per-tag precision, recall and F-score
of all systems. 
Table~\ref{baselinetable0} shows good FoM 
of our three systems, 
which are on-par with those of four other state-of-the-art approaches
(included in the table).
We also test all systems on the three datasets,
restricted to the tag vocabulary of Vocals and Non-Vocals.
Table~\ref{baselinetable2} shows very good 
results for these systems. 
%

\begin{table}[h] 
\centering
\begin{tabular}{|l|ccc|}
\hline
& \multicolumn{3}{c|}{CAL500 (97 tags)}  \\
& P & R & F \\
\hline
SVMBFFs  & 0.40 & 0.40 &  0.40 \\
VQMM &  0.38 &  0.46 &  0.42\\
SRCAM &  0.34 & 0.57 & 0.42 \\
HEM-DTM \cite{Coviello2012} &  0.45 & 0.22  & 0.26  \\
\cite{Miotto2010b} &  0.44 & 0.23  & 0.30  \\
\cite{Xie2011} &  0.45 & 0.23  & 0.30  \\
\cite{Nam2012b} &  0.48 & 0.26  & 0.29  \\
\hline
\end{tabular}
\caption{Average per-tag  precision, recall and F-score
 of the three systems, compared to recent systems, on 
CAL500 restricted to the 97 most frequent tags, 5-fold cross-validation procedure. 
} 
\label{baselinetable0}
\end{table}

\begin{table*}[t]\scriptsize
\centering
\begin{tabular}{|ll|ccc|}
\hline
& & \multicolumn{3}{c|}{CAL500}  \\
& & P & R & F \\
\hline
\multirow{2}{*}{$S_1$} & V  & $0.92\pm0.02$ & $0.99\pm0.00$ & $0.95\pm0.01$  \\
& NV & $0.78\pm0.04$ & $0.33\pm0.17$ &  $0.45\pm0.18$ \\ \hline
\multirow{2}{*}{$S_2$} & V &  $0.93\pm0.01$ &  $0.96\pm0.02$ &  $0.95\pm0.01$\\
& NV & $0.63\pm0.11$ & $0.48\pm0.09$ &  $0.54\pm0.02$ \\ \hline
\multirow{2}{*}{$S_3$} & V &  $0.94 \pm 0.01$ &  $0.95 \pm 0.02$ &  $0.95 \pm 0.01$ \\
& NV & $0.60 \pm 0.12$ &  $0.55 \pm 0.05$ &  $0.57 \pm 0.08$ \\
\hline
\end{tabular}
\hspace{-.2cm}
\begin{tabular}{ccc|}
\hline
\multicolumn{3}{c|}{MagTag5k} \\
 P & R & F \\
\hline
$0.88\pm0.01$ & $0.91\pm0.02$ & $0.89\pm0.01$ \\
$0.79\pm0.03$ & $0.72\pm0.02$ & $0.75\pm0.01$ \\ \hline
$0.85\pm0.02$ & $0.85\pm0.03$ & $0.85\pm0.01$\\
$0.66\pm0.02$ & $0.67\pm0.06$ & $0.66\pm0.02$\\ \hline
$0.88 \pm 0.01$ &  $0.92 \pm 0.02$ &  $0.90 \pm 0.004$ \\
$0.80 \pm 0.03$ &  $0.73 \pm 0.04$ &  $0.76 \pm 0.01$ \\
\hline
 \end{tabular}
\hspace{-.2cm}
\begin{tabular}{ccc|}
\hline
\multicolumn{3}{c|}{MSD24k} \\
 P & R & F \\
\hline
$0.89\pm0.01$ & $0.92\pm0.01$ & $0.91\pm0.01$ \\
$0.82\pm0.02$ & $0.77\pm0.03$ & $0.80\pm0.02$ \\ \hline
$0.85\pm0.01$ & $0.80\pm0.00$ & $0.83\pm0.01$\\
$0.62\pm0.01$ & $0.71\pm0.03$ & $0.66\pm0.02$ \\ \hline
$0.89 \pm 0.01$ &  $0.94 \pm 0.01$ &  $0.91 \pm 0.004$ \\
$0.86 \pm 0.01$ & $0.74 \pm 0.03$ &  $0.80 \pm 0.02$ \\
\hline
\end{tabular}
\caption{
Average $\pm$ standard deviation, for Precision, Recall and F-Score 
for the 3 systems on CAL500, MagTag5k and MSD24k (respectively with 2-fold, 3-fold and 3-fold cross-validations). 
Vocabulary restricted to Vocals (``V'' rows) and Non-Vocals~(``NV'' rows) .
$S_1$ is SVMBFFs,
$S_2$ is VQMM, and
$S_3$ is SRCAM. 
} 
\label{baselinetable2}
\end{table*}

\begin{table*}[t]\scriptsize
\centering
\begin{tabular}{|lc|cc|}
\hline
& & \multicolumn{2}{c|}{CAL500}  \\
& & Fold 1 & Fold 2  \\
\hline
\multirow{2}{*}{$S_1$} & $\mathcal{F}_{def}(\Phi)$  & $\surd$ & $\surd$ \\
& $\mathcal{F}_{inf}(\Phi)$ & $1.0$ & $1.0$ \\ \hline
\multirow{2}{*}{$S_2$} & $\mathcal{F}_{def}(\Phi)$ &  $\bm{\surd}$ &   $\surd$ \\
& $\mathcal{F}_{inf}(\Phi)$ & $\bm{0.95}$ & $0.97$\\ \hline
\multirow{2}{*}{$S_3$} & $\mathcal{F}_{def}(\Phi)$ &  $\surd$ &  $\surd$  \\
& $\mathcal{F}_{inf}(\Phi)$ &  $0.98$ &  $0.97$ \\
\hline
\end{tabular}
\hspace{-.2cm}
\begin{tabular}{ccc|}
\hline
\multicolumn{3}{c|}{MagTag5k} \\
 Fold 1 & Fold 2 & Fold 3 \\
\hline
 $\surd$ & $\surd$ &  $\surd$ \\
$0.89$ & $0.96$ & $0.89$ \\ \hline
$\bm{\surd}$ & $\surd$ & $\surd$\\
$\bm{0.97}$ & $0.97$ & $0.98$\\ \hline
$\surd$ &  $\surd$ & $\surd$ \\
$0.98$ & $0.99$ &  $0.99$ \\
\hline
 \end{tabular}
\hspace{-.2cm}
\begin{tabular}{ccc|}
\hline
\multicolumn{3}{c|}{MSD24k} \\
 Fold 1 & Fold 2 & Fold 3 \\
\hline
$\surd$ & $\surd$ & $\surd$ \\
$0.99$ &  $0.93$ & $0.93$ \\ \hline
$\bm{\surd}$ & $\surd$ & $\surd$ \\
 $\bm{0.96}$ &  $0.95$ &  $0.95$ \\ \hline
$\surd$ & $\surd$ & $\surd$ \\
$0.99$ & $0.99$ & $0.99$ \\
\hline
\end{tabular}
\caption{
Effect of the deflation and inflation
procedures applied to test sets. 
$S_1$ is SVMBFFs,
$S_2$ is VQMM, and
$S_3$ is SRCAM.
Columns correspond to the test folds (corresponding training data are the remaining folds). 
$\surd$ denotes cases where 
a system 
with initial  performance superior to random ($p < \alpha = 0.01$ in~(\ref{eq:pvalueexp1}))
performs consistently to random  after
deflation  of the test set.
Reported average per-tag F-scores after inflation of the test sets
($\mathcal{F}_{inf}(\Phi)$ rows) are close to perfect.
In bold, results obtained with data which train/test divergence is reported in the second column of Table~\ref{tab:div}.
}
\label{filteringtable2}
\end{table*}

\subsection{On absolute performance (tasks A1 and A2 in practice)}\label{experiment1}

We now perform tasks A1 and A2 using the methodology in Section~\ref{sec:Methodology}.
For a given system $S$ (which is already trained on a subset of data folds) 
and a test dataset $\Phi$ (remaining fold of dataset), 
we aim to find the set of irrelevant transformations $\mathcal{F}_{def}(\Phi)$ (for ``deflation'') and $\mathcal{F}_{inf}(\Phi)$  (for ``inflation'')
such that $S$ performs no better than random for $\mathcal{F}_{def}(\Phi)$,
and $S$ performs close to perfectly for $\mathcal{F}_{inf}(\Phi)$.
%
Section~\ref{irrelevance} below confirms the irrelevance of our transformations using covariate shit and listening tests.

\begin{figure}[t]
\centering
\includegraphics[width=0.8\columnwidth]{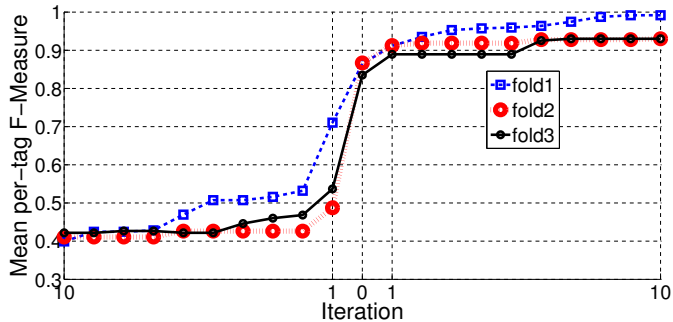}
\caption{Mean per-tag F-measure (average over Vocals and Non-Vocals) with respect to 
ten successive iterations of the 
deflation
 procedure (iterations left to the origin)
and 
inflation
 procedure (iterations right to the origin), as
detailed in Section~\ref{sec:Methodology},
for three SVMBFFs systems tested on three different folds of MSD24k.
F-measure at iteration $0$ for the three folds ($\approx 0.85$) corresponds to 
average performance of SVMBFFs on MSD24k as can be seen on Table~\ref{baselinetable2}.
}
\label{fig:iterations}
\end{figure}

Figure~\ref{fig:iterations} shows the FoM of three SVMBFFs systems,
trained on three combinations of two MSD24k folds and tested on the three 
respectively remaining folds. FoM is plotted versus iterations of the 
deflation and inflation procedures applied to the test set.
On all three folds, we see that our procedures yield clear decrease and increase in
FoM in very few iterations.

\begin{figure}[t]
\centering
\includegraphics[width=0.8\columnwidth]{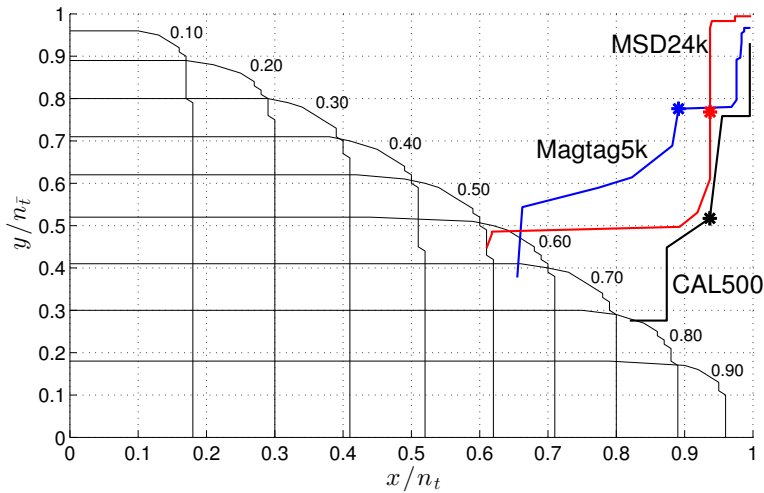} 
\caption{For three systems created using the SRCAM approach, 
we are able to transform the test data --CAL500 (black), MagTag5k (blue), and MSD24k (red)--
such that their performance is 
near perfect ($\mathcal{F}_{inf}(\Phi)$, top right corner),
or consistent 
with 
that expected from a random system $R(p_t)$
($\mathcal{F}_{def}(\Phi)$, within thin black lines, where \(p > \alpha = 0.01\))
that randomly picks $t$ with probability $p_t$
(illustrated here between 0.10 and 0.90, in steps of 0.10)
and $\bar t$ with probability $1-p_t$.
Each star marks the ``starting position'' of the system.
$x/n_t$ is the ratio of correctly classified instances of Vocals, 
$y/n_{\bar t}$ is the ratio of correctly classified instances of Non-Vocals.
}
\label{fig:SRCAMdecay}
\end{figure}

Figure~\ref{fig:SRCAMdecay} shows the FoM
of three SRCAM systems trained on one CAL500 fold (black line), 
two MagTag5k folds (blue line) and two MSD24k folds (red line) respectively, and tested 
on the remaining fold of the respective dataset. 
The line corresponding to each system represents change in FoM
with respect to successive transformations of the test set. 
In other words, the opposite ends of a given line correspond to the FoM
obtained either after deflation or inflation of the test set.
One can see that the performance of all systems
can take on drastically different values
after few iterations of irrelevant transformations. Namely, the performance of each system can be significantly better than random 
(outside region demarcated by black lines),
to no better than random (inside region, according to~(\ref{eq:pvalueexp1})).

Table~\ref{filteringtable2} reports results for all systems using SVMBFFs, VQMM and SRCAM approaches, 
on all folds of the three datasets.
Each cell in the table corresponds to a system built using one of the three approaches,
trained on some data folds of a given dataset, 
and tested on the remaining fold.
Results correspond to either the deflation or inflation procedures.
The performance of each system can vary between almost perfect 
to no better than random, while the diversity of experimental conditions has no effect
on whether a given piece of music includes singing voice or not, and is perceived as such.

\subsection{On relative performance (tasks B1 and B2 in practice)}\label{experiment2}

We now perform tasks B1 and B2 
using the methodology in Section~\ref{sec:Methodology}.
For two given systems $S_i$ and $S_j$ 
(already trained on a subset of data folds)
and a test dataset $\Phi$ 
(remaining fold),
we aim to find a transformation $\mathcal{F}_i$
such that $S_i$ performs significantly better (according to (\ref{eq:pvalueexpB1})) than $S_j$
on $\mathcal{F}_i(\Phi)$, and another transformation $\mathcal{F}_j$
such that the opposite is true
on $\mathcal{F}_j (\Phi)$.

After conducting experiments for all possible pairwise comparisons of any two systems among 
SVMBFFs, VQMM, and SRCAM, on any possible test set among each of the three datasets 
we use, we can report that it is always possible, in a few iterations, to find an irrelevant transformation of 
any test set so that any two systems are alternatively the best.\footnote{See the article's companion webpage (link in in Section~\ref{intro}) 
for results and their reprocuction  (i.e. $3$ systems $*$ 2 conditions $*$ (2+3+3) folds $=48$  comparisons in total).}

\subsection{Testing the irrelevance of the transformations}\label{irrelevance}

\subsubsection{On the irrelevance of the transformations with respect to covariate shift}\label{featureirrelevance}

In our experimental procedure, measuring covariate shift is important for verifying irrelevance of the transformations.
We need to make sure that there is no significant divergence between the feature distributions
of train and test data.
For this, we follow the method proposed by 
\cite{Ben-David}. 
They show that an upper bound on the divergence $d_{\cal H}({\cal D}, {\cal D'})$ between two distributions ${\cal D}$ and ${\cal D}'$
can be estimated from an empirical divergence $\hat d_{\cal H}({\cal U}, {\cal U}')$ 
computed from finite samples ${\cal U}$ and ${\cal U}'$ of these distributions.

The method for computing $\hat d_{\cal H}({\cal U}, {\cal U}')$
consists in labelling each instance $x \in {\cal U}$ with~0, and each instance $x \in {\cal U'}$ with~1.
Then training classifiers\footnote{${\cal H}$ is a class of functions from  features to tag,
which, for consistency with the rest of this article, we refer to as a set of classifiers (e.g. linear perceptrons). 
The correct naming would be a ``hypothesis class''~\cite{Ben-David}.} 
$h \in {\cal H}$ to discriminate between instances of ${\cal U}$ and ${\cal U}'$.
In a testing phase, one can then compute a confusion matrix for each classifier 
$h$ and compute $\hat d_{\cal H}({\cal U}, {\cal U}')$ as follows (lemma~2 in~\cite{Ben-David}):
\begin{equation}
\hat d_{\cal H}({\cal U}, {\cal U}') = 2 \biggl(1 - \min_{h\in{\cal
  H}}\biggl[\frac{1}{m} \sum_{x:h(x) = 0} I[x \in {\cal U}] +  
  \frac{1}{m} \sum_{x:h(x)=1} I[x \in {\cal U}']\biggr]\biggr)
\label{eq:lemma2}
\end{equation}
where $m$ is the number of instances in ${\cal U}$ and ${\cal U}'$ and $I[x]$
indicates class membership of $x$ (i.e. $I[x \in {\cal U}] = 1$ if $x \in {\cal U}$).
Smaller values in~(\ref{eq:lemma2}) refer to smaller divergence.
As noted in~\cite{Ben-David}, it is not feasible to compute~(\ref{eq:lemma2}) with the minimum 
over \emph{all possible} classifiers $h \in {\cal H}$.
In our experiments below, we therefore compute the minimum over ten different classifiers (which
we choose to be linear perceptrons).


An upper bound on $d_{\cal H}({\cal D}, {\cal D'})$ is then given by the following equation (lemma~1 in~\cite{Ben-David}):     
\begin{equation}
d_{\cal H}({\cal D}, {\cal D'}) \leq {\hat d}_{\cal H}({\cal U}, {\cal U'}) + 4 \sqrt{\frac{d \log(2m) + \log(2/\delta)}{m}}
\label{eq:lemma1}
\end{equation}
where $d$ is ${\cal H}$'s VC dimension~\cite{Ben-David}, 
and $\delta \in (0,1)$ is a confidence parameter.

In the case where the samples ${\cal U}$ and ${\cal U}'$ are drawn from the same distribution,
for instance if ${\cal U}$ is a sample of a training fold $\Psi$ and ${\cal U}'$ a sample of a test fold $\Phi$
of the same dataset,
the classifiers $h$ should do a bad job a discriminating between instances of ${\cal U}$ and ${\cal U}'$.
$d_{\cal H}({\cal D}, {\cal D'})$ should therefore be low. 
In our experiments below, we precisely compare the divergence
in such cases (namely when no data is transformed) to the divergence when some data is transformed
by inflation or deflation.

The first column of Table~\ref{tab:div} corresponds to cases where
we define ${\cal U}$ as 100k randomly selected frames from one data fold of a given dataset,
and ${\cal U}'$ as 100k randomly selected frames of the complementing fold(s) of that dataset.\footnote{Recall that 
for computing~(\ref{eq:lemma2}), the labelling of instances $x \in {\cal U}$ with 0 and $x \in {\cal U'}$ with 1
have nothing to do with Vocals and Non-Vocals tags. ${\cal U}$ and ${\cal U}'$ are random frames from Vocals and Non-Vocals instances.}
We then use half of ${\cal U}$ and half of ${\cal U}'$ for training simple linear perceptrons, and the remaining 
halves for computing~(\ref{eq:lemma2}). 
Two trials were done for each dataset. 
In these cases, in the first column of Table~\ref{tab:div}, in each line, the data is coming from a single dataset,
and \emph{no} instance is transformed, 
the divergence values obtained are therefore representative of standard cases of autotagging evaluation (i.e. cross-validation)
where one can consider that there is no significant divergence in feature distributions of 
train and test data, i.e. no covariate shift. The inter-row differences provide
examples of non-significant variability in the computation of the divergence.\footnote{Divergence upper bounds are $\neq 0$
because of the 
second term in the right-hand side of~(\ref{eq:lemma1}) and by
the fact that a linear perceptron is a weak classifier. A better
classifier would probably give tighter bounds. }

The second column of Table~\ref{tab:div} corresponds to cases where
we define ${\cal U}'$ as 100k randomly selected frames of the \emph{transformed} fold of a given dataset 
(namely the transformed fold used for test in inflation and deflation experiments which results are 
reported in bold in Table~\ref{filteringtable2}),
and where we define ${\cal U}$ as 100k randomly selected frames from the complementing data fold(s) of that dataset.
%
The second column shows that when applying transformations (either inflation or deflation) to the test set,
the upper bounds for the divergence  between training and test sets are relatively low,
and sensibly the same as when no transformation is applied (i.e., in the first column).
This provides evidence of the irrelevance of the transformations with respect to covariate shift.


\begin{table}[h]
\centering
\begin{tabular}{|l|lc|lc|}
\hline
& & $\Psi$ \textit{vs} $\Phi$  & & $\Psi$ \textit{vs} $\mathcal{F}(\Phi)$   \\ \hline
\multirow{2}{*}{CAL500} & trial 1  & 0.34 & $\mathcal{F}_{inf}(\Phi)$ & 0.35  \\
 & trial 2 & 0.38 &        $\mathcal{F}_{def}(\Phi)$ &              0.39 \\ \hline
\multirow{2}{*}{MagTag5k} & trial 1  & 0.40 & $\mathcal{F}_{inf}(\Phi)$ & 0.34  \\
 & trial 2 & 0.37 &         $\mathcal{F}_{def}(\Phi)$ &             0.36\\ \hline
\multirow{2}{*}{MSD24k} & trial 1  & 0.24 &         $\mathcal{F}_{inf}(\Phi)$  &             0.26  \\
 & trial 2 & 0.27 &      $\mathcal{F}_{def}(\Phi)$ &                0.39 \\ \hline
\end{tabular}
\caption{\small{Upper bounds for $d_{\cal H}(\cal D, \cal D')$,
computed as (\ref{eq:lemma1}).
$\mathcal{F}_{inf}(\Phi)$ and $\mathcal{F}_{def}(\Phi)$ rows correspond to inflation or deflation procedures applied to the test set
which corresponding performances are reported in bold in Table~\ref{filteringtable2}. 
}
\label{tab:div}}
\end{table}

\subsubsection{On the perceptual irrelevance of the transformations}\label{perceptualirrelevance}

A key aspect in our experiments relies on our assumption of perceptual irrelevance 
of the deflation and inflation procedures.
In order to verify this assumption,
%
we perform a listening test,
where 152 subjects are asked to rate 32 audio stimuli
 with respect to whether they contain singing voice or not.
Stimuli are representative of those used in experiments 
with autotagging systems in Sections~\ref{experiment1} and~\ref{experiment2},
i.e. half of the stimuli are ``originals'', while the other half are transformed according to 
deflation or inflation procedures. 
Results show that recognition of singing voice is very good, i.e. $\approx 98\%$, 
and that there is no significant effect of the condition (original or transformed).
More details  are available in 
Appendix~B.

\vspace{-0.1cm}


\section{Summary and Discussion}\label{discussion}

In this article, we tackle the issue of validity in the evaluation of music autotagging systems.
For a given music autotagging system, a valid evaluation means that
there is a 
high positive correlation between
its figure of merit and its true performance on the task for which it has been designed.
This is essential for making relevant conclusions about a system's
performance in laboratory conditions (and all the more in real-world conditions).
Validity is, more generally, paramount to guarantee continued improvements in autotagging system research and development.
Our main contributions in this paper are the formalization of the notion of validity in autotagging evaluation and
the proposal of a method for testing it (with available code), which centers on the control of experimental conditions
via irrelevant transformations of input signals.

We demonstrate the use of our method with three autotagging systems
in a simple two-class setting (i.e. recognizing the presence or absence of singing voice in an excerpt).
We find we can make all three perform as well or as poorly as we like by irrelevant transformations.
Although these systems initially appear to be on-par with current
state-of-the-art, their FoM do not provide valid indicators of their
true performance on the task of recognizing the presence or absence of singing voice in an excerpt,
and do not provide valid indicators for comparing them in that task.

An important point to clarify is that
our method does not aim to answer questions regarding system performance in the real world.
It is designed first and foremost to answer questions about what the systems
have learned to do. And our conclusions are limited to particular datasets.
In other words, our experiments aim to answer
whether the observation of the systems' FoM, or comparisons thereof,
warrant any conclusion about the actual capacity of these systems to annotate CAL500, MagTag5k, or MSD24k
data with the concept of singing voice.
%
%
We claim that our experiments provide evidence that this is in fact not the case.
Questioning whether these systems would be able to apply that
concept  
in the real world
(where e.g. covariate shift would probably happen) is another question altogether,
which we do not address in this article.

Since we consider a special case of autotagging 
that is simpler than the general case of multi-label classification, 
i.e., we consider only music labeled using two mutually exclusive tags, ``Vocals'' and ``Non-Vocals'',
the generality of our work here may appear limited;
the autotagging systems used in this article are indeed not designed
only for this two-class problem, but for multi-label classification
(including these two classes nevertheless).
We also do not claim that the evaluation of these systems
is necessarily uninformative for any possible tag.
Instead, we just show that even for what should
be a simple case for these systems, it is 
not possible to conclude upon the degree to which they have learned to perform the task.
We do claim that this sheds doubt on knowledge we could obtain with certitude in more 
difficult cases.
For instance, if we cannot make valid conclusions about these systems' ability
to recognize singing voice, how could these evaluation approaches suddenly 
serve for solid conclusions on the finer, and more subjective tags like
``Vocals-Aggressive,'' ``Vocals-Call \& Response,'' ``Vocals-Falsetto,''
and ``Vocals-Rapping''? 

It is important to clarify that, although our method uses signal transformations at its core, 
it is fundamentally different from robustness testing.
We ask a different scientific question.
While robustness testing asks ``How does the performance of $S$ change in condition X?'', 
we ask ``Does the evaluation of $S$ provide a valid indication of its true performance?''
%
%
%
More than testing the robustness of a particular autotagging \emph{system},
our claims in this article are relative to the validity of the \emph{evaluation} itself.
In other words, we use a similar machinery as robustness tests, but only as 
part of 
a method whose aim is to test evaluation validity.
Further, in existing work on robustness testing~\cite{Sigurdsson,JensenCEJ09,UrbanoISMIR2014,journals/taslp/GouyonKDATUC06,MauchE13}, 
experimental conditions are made increasingly more challenging,
and decreasing performance
is assumed to illustrate disruptibility of a system and its inability to 
complete its task
\emph{under all possible conditions}. 
Robustness testing is thought to highlight e.g. possibly overestimated FoM,
but representative FoM nevertheless.
Thus the comparison and ranking
of several systems is still thought to be possible and informative.
In contrast, we claim that the diverse experimental conditions
(i.e. all  possible $\mathcal F(\Phi)$, including no transformation at all) 
should not reflect significantly on the behavior of systems if they are pairing 
audio signals with tags in a meaningful way.
%
Under 
these experimental conditions, 
we showed that not only the estimated performances of three systems can drop to random,
but it can also ascend to almost perfect, thus providing no valid indication of true performance of these systems 
on a simple task, and hence uninformative with regards to these systems' ranking.

The erratic behavior of systems' FoM
under our experimental conditions
does not mean that the performance measure itself (e.g. the average per-tag F-score) is to blame,
or that the systems we consider are unable to learn from data.
Instead, it may indicate that
what the systems are learning may not necessarily be
what they are assumed to have learnt, i.e. the particular dimensions of interest
to the evaluator (e.g. the presence or absence of singing voice).
Observing correlations between  some characteristics of music audio signals and 
a particular tag cannot by itself lead to the conclusion that the former are necessarily relevant to the latter. 
Such correlations are just an indication that the former \emph{may} be relevant to the latter~\cite{citeulike:5820221}.
In other words, irrelevant characteristics may be confounded with the dimensions of interest~\cite{Sturm2014a}.
Indeed it is likely that the autotagging systems we consider are able to learn from training data
an uncontrolled (and unidentified) confounding variable, rather than the presence or absence of singing voice.
This factor is highly correlated with the presence/absence of singing voice
on the datasets we considered, hence explaining the good FoM  in Table~\ref{baselinetable2}. 
(Note that a similar argument on the impact of confounding variables on estimated performance
was made in previous MIR work, in the particular case of artist and album effects~\cite{Pampalk2005,Flexer2007}.)
Although our transformations are irrelevant to singing voice,
they do affect that confounding variable, hence explaining the large variations in FoM we see e.g. in Table~\ref{filteringtable2}. 
If, for instance, all excerpts tagged ``Vocals'' in a dataset are loud, and all excerpts tagged ``Non-Vocals'' are quiet,
then the evaluation of a system exploiting only loudness to discriminate between the two
will measure the system to be perfect, yet providing no validity for drawing reasonable conclusions
on the true performance of that system for actually recognizing~singing~voice in that~dataset.

How could one reliably conclude anything about the ability 
of a given autotagging system to perform the task at hand?
Before being a question of which statistical test to use,
or which figures of merit to avoid, it is first and foremost
a matter of the design, implementation, and analysis of an evaluation that
is valid with respect to estimating true performance.
An evaluation is either valid or invalid with respect to the question
one is attempting to address --no matter the actual results of the evaluation.
\cite{UrbanoJIIS2013} 
discuss several important notions of validity
in scientific experiments, and how they relate to MIR.
Another critical component is formalizing evaluation~\cite{Bailey,BobCMMR2014}.
In this paper we build on previous research by proposing a method (and code) for testing validity in music autotagging 
experiments, adapting the method in~\cite{Sturm2014a}, which is reproduced independently in~\cite{Szegedy2014} for image tagging.

Another important point to reiterate here is that what is general in our proposed method for evaluation validity
is the notion of ``irrelevant transformation,'' not the particular transformation itself (i.e. our time-invariant filtering).
Indeed, the irrelevance of a particular transformation largely depends on the task at hand.
In this article, for the purpose of demonstrating the use of our method, we show that
our time-invariant filtering is irrelevant to the specific task of Vocals/Non-Vocals autotagging.
Time-stretching, e.g.,  may have been another option for that task~\cite{Sturm2014}.
On the other hand, time-invariant filtering would probably not be appropriate
to our method if the task at hand were to annotate music audio signals 
with tags related e.g. to audio production quality, such as  ``low-fi'' vs. ``hi-fi'' for instance.
In other words, future extensions of the work presenting here
may call for different transformations.

Future work will look into which other irrelevant transformations can
be designed for testing the validity of evaluation in other MIR tasks.
We believe that building our method into MIREX-like campaigns would also be of interest.
\cite{Bailey} provides a very interesting starting point
to further work on the formalization of the notion of confounds
in MIR research.
Another interesting avenue for future work is the adaptation to music autotagging of existing research
on the design of systems that can be used in different conditions than those under which they were 
developed.
For instance, an adaptation of our method may be used to attempt to train better systems, as suggested in~\cite{Szegedy2014}. 
Namely, one could train systems on datasets ``enriched'' by carefully designed perturbations of instances.
Other methods to train systems able to cope with different conditions than those under which they were developed
may be adapted from~\cite{Quionero-Candela:2009:DSM:1462129,PanY09TKDE,Ben-David,Sugiyama2007}.

\vspace{-0.3cm}

\section*{Acknowledgments}

FG (\url{fgouyon@inesctec.pt}) and NH are with INESC TEC, Porto, Portugal.
BLS is with the Audio Analysis Lab, AD:MT, Aalborg University
Copenhagen, 
Denmark.
JLO is with INESC TEC and FEUP, Porto, Portugal.
TL is with the Science Faculty of Lisbon University, Portugal.
FG acknowledges the support of 
the Media Arts and Technologies project (MAT), NORTE-07-0124-FEDER-000061, 
financed by the North Portugal Regional Operational Programme (ON.2 Ð O Novo Norte), 
under the National Strategic Reference Framework (NSRF), through the European Regional Development Fund (ERDF), 
and by national funds through the Portuguese funding agency Funda\c{c}\~{a}o para a Ci\^{e}ncia e a Tecnologia (FCT). 
%
%
BLS acknowledges the support of Independent Postdoc Grant 11-105218 from Det Frie Forskningsr\aa d. 
We thank Matthew Davies, Marcelo Caetano, Jo\~{a}o Gama, Guy Madison, Doug Turnbull 
and anonymous reviewers for useful discussions.

\bibliographystyle{apalike}
\bibliography{tagging}

\section*{Appendices}

\subsection*{A --- Tags  defining ``Vocals'' in CAL500 and MSD24k}\label{listings}

\begin{lstlisting}[style=tags, caption={CAL500 tags for tag Vocals}]
Instrument_-_Backing_vocals
Instrument_-_Female_Lead_Vocals
Instrument_-_Male_Lead_Vocals
Vocals-Aggressive
Vocals-Altered_with_Effects
Vocals-Breathy
Vocals-Call_&_Response
Vocals-Duet
Vocals-Emotional
Vocals-Falsetto
Vocals-Gravelly
Vocals-High-pitched
Vocals-Low-pitched
Vocals-Monotone
Vocals-Rapping
Vocals-Screaming
Vocals-Spoken
Vocals-Strong
Vocals-Vocal_Harmonies
Instrument_-_Female_Lead_Vocals-Solo
Instrument_-_Male_Lead_Vocals-Solo
\end{lstlisting}

\begin{lstlisting}[ style=tags, caption={MSD24k tags for tag Vocals.}]
a_breathy_male_lead_vocalist
a_distinctive_male_lead_vocal
a_dynamic_female_vocalist
a_dynamic_male_vocalist
a_female_vocal
a_gravelly_male_vocalist
a_laid_back_female_vocal
a_smooth_female_lead_vocal
a_smooth_male_lead_vocalist
a_vocal-centric_aesthetic
an_aggressive_male_vocalist
an_emotional_female_lead_vocal_performance
an_emotional_male_lead_vocal_performance
jazz_vocals
\end{lstlisting}

\vspace{-0.2cm}

\subsection*{B --- Listening test}\label{perceptualtest}



The listening test includes 32 stimuli of 30~s each (8 stimuli with singing voice, 8 without, 
and their 16 transformed versions).
%
The stimuli and one test sound example are normalized with respect to loudness.
%
%
The listening test was performed online via a web-based questionnaire, written in English.
The questionnaire was available online between 15th July-2nd August 2013.
Few participants reported sound playback issues, 
consequently their responses were not included in the analyses.

Before proceeding to the experiments, participants were asked to set up the volume 
to a comfortable level by listening to a test sound example (not included in the stimuli).
Each participant listened to the 32 stimuli
and was asked to rate whether yes or no it contained a singing voice.
An entire session took between 16-20~min to complete.
By listening to  
the full list of stimuli,
participants rated both conditions (original and transformed) of each stimuli.
In order to control for a potential bias in ratings of the second condition heard,
that would result from having previously heard the other condition,
participants were assigned to one of 2 groups corresponding to a difference in presentation order: 
group A listened to the 16 original stimuli first and then to the 16 transformed stimuli, 
while group B did the opposite.
Within each 16-stimuli block, the ordering of stimuli was done randomly on a subject-by-subject basis.
Subjects were attributed group A or B in an alternate fashion.
Participants could listen to each stimulus only once, and they had to listen to the
full duration of the stimuli before being able to listen to the next one.


A total of 254 participants took the test, of which 152 fully completed the test (79 men, 73 women, average age~$\pm$~$\sigma=~25.3y\pm6.3$).
The participants were recruited via emails, sent to
international mailing lists.
Participants were not paid.
The following analyses are based 
on the 152 complete responses.
There are 76 participants in both groups A and B.


Overall, the recognition of the presence of singing voice was very good, i.e. 98.1\%$\pm$1.6.
Considering all different conditions (original stimuli, transformed stimuli, group A, group B),
and all combinations of conditions, correct recognition rates range between 97-99\%.

One might raise 
the question whether 
listening to the same original and transformed stimuli successively
might have implied a bias in recognition rates, i.e.
artificially higher recognition rates for transformed stimuli for participants of group A, and inversely,
higher recognition rates for original stimuli for participants of group B.
A paired \emph{t}-test was therefore conducted to compare recognition rates of singing voice presence
for group A in original vs. transformed stimuli conditions.
There was no significant difference in the recognition rates for original ($M=97.5\%$, $SD=2.5$)
and transformed conditions ($M=98.0\%$, $SD=2.0$);
 $t(15)=-1.19$, $p=0.25$.
A similar test was conducted 
for group B. 
Here also, there was no significant difference in the recognition rates for transformed ($M=98.4\%$, $SD=1.3$)
and original conditions ($M=98.3\%$, $SD=1.9$);
 $t(15)=0.25$, $p=0.80$.
These results suggest that  listening to the two conditions in a row did not
imply a bias in participants recognition rates.
Which therefore leads us to validate our chosen experimental design and
to use the full amount of data collected in further analyses.


We performed a two-way ANOVA in order to determine whether  (i) 
the presentation order (i.e. original version first, or alternatively transformed versions first)
and, most importantly, (ii) 
the stimuli condition (original vs. transformed), 
had an effect on correct recognition of singing voice in stimuli.

The results showed 
no significant effect of the presentation order ($F(1,60)=1.59$, $p=0.21$), hence
corroborating results reported above,
and 
no significant effect of the stimuli condition ($F(1,60)=0.35$, $p=0.56$).
We also found that the interaction effect between condition and presentation order was non-significant
($F(1,60)=0.17$, $p=0.68$).

These results indicate that the transformation procedures 
do not appear to have any noticeable effect on human perception of
the presence/absence of singing voice. 


\end{document}